\documentstyle[amssymb,aps,epsfig,float,twocolumn]{revtex}
\restylefloat{figure}
\title{Lattice polaron formation: 
Effects of non-screened electron-phonon interaction}
\author{H.~Fehske$^a$, J. Loos$^b$, and G.~Wellein$^c$ }
\address{$^a$Physikalisches Institut, Universit\"a{}t Bayreuth, 
  95440 Bayreuth, Germany\\
  $^b$Institute of Physics, Czech Academy of Sciences, 16200
Prague, Czech Republic\\
  $^c$Regionales Rechenzentrum Erlangen, Universit\"a{}t Erlangen, 
  91058 Erlangen, Germany\\ 
{\rm (\today)}
  }
\address{~\parbox{14cm}{\rm
    \medskip
We explore the quasiparticle properties of lattice polarons on the basis
of a quite general electron--phonon Hamiltonian with a long--range
displacement--type of interaction. To treat the dynamical quantum phonons 
without significant loss of accuracy we adapt an exact 
Lanczos diagonalization method and compute various static 
and dynamical quantities, such as the electron-lattice correlation function, 
the polaron band dispersion, the effective polaron mass, the kinetic energy,  
the single--particle spectral function, and the optical conductivity, 
on finite one--dimensional lattices for a wide range of model parameters. 
We compare the results with those obtained for the standard 
Holstein model with short-range electron-phonon interaction only. 
    }}
\def\ep{\varepsilon_p}
\def\eps{\tilde{\varepsilon}_p}

\def\ts{\tilde{t}}
\def\tsq{\tilde{t}^{\,2}}
\def\vk{\varkappa}
\def\o{\omega}
\def\ho{\omega_0}
\def\mP{{\mit \Phi}}
\def\mS{{\mit \Sigma}}
\def\cC{{\cal{C}}}
\def\cD{{\cal{D}}}
\def\cG{{\cal{G}}}
\def\cH{{\cal{H}}}
\def\cS{{\cal{S}}}
\def\cT{{\cal{T}}}
\def\cU{{\cal{U}}}
\begin{document}
\maketitle
\section{Introduction}
The classical polaron problem~\cite{La33} has received renewed attention on 
account of the observation of polaronic effects in several 
important classes of materials, including high-temperature cuprate 
superconductors and colossal magneto--resistance 
manganites~\cite{SAL95,JHSRDE97}.
Remarkably even the much simpler case of free electrons interacting 
with optical phonons in ionic crystals is still not completely 
understood. From a theoretical point of view the challenge is to 
describe the crossover from an only weakly dressed charge carrier to the 
strongly mass-enhanced, i.e., less mobile, polaronic quasiparticle 
with increasing electron--lattice coupling  strength. 
Depending on the relative importance of the short-- or 
long--range electron--phonon (EP) coupling simplified models 
of Holstein~\cite{Ho59a} or Fr\"ohlich~\cite{Fr54} type, 
respectively, have been studied over the last five decades. 
However, despite extensive analytical work, in the physically most 
interesting crossover regime, up to now, the only reliable 
results came from numerical studies, 
such as finite--cluster exact diagonalizations 
(ED)~\cite{RT92,Ma93,AKR94,St96,WF97,FLW97}, 
(Quantum) Monte Carlo (QMC) simulations~\cite{RL82,BVL95}, 
density--matrix renormalization--group (DMRG) 
approaches~\cite{JW98b,ZJW98}, and global-local~\cite{RBL98} 
or variational methods~\cite{BTB99}.   

Recently the formation of small polarons was investigated 
by Alexandrov and Kornilovitch~\cite{AK99} applying
a new path-integral Monte-Carlo algorithm~\cite{Ko98,Ko99}. 
These authors introduced the following EP Hamiltonian 
\begin{eqnarray}
\cH&=&-t\sum_{\langle j,j'\rangle} c_{j'}^{\dagger} c_{j}^{}
+\ho\sum_l  ( b_l^{\dagger} b_l^{}+ 
\mbox{\small $\frac{1}{2}$})\nonumber\\ 
&&-\sum_{j,l} f_l(j)c_{j}^{\dagger} c_{j}^{}
x_0 ( b_l^{\dagger}  + b_l^{}) \,.
\label{ED1}
\end{eqnarray}
Here $c_j^{[\dagger]}$ and $ b_j^{[\dagger]}$ denote fermionic 
and bosonic annihilation [creation] operators, respectively. 
Restricting ourselves to the one--dimensional (1D) case, 
$\cH$ describes an electron in a Wannier state on site $j$ of 
an infinite chain which interacts with the vibrations  
of all ions of another chain 
via a ``density-displacement'' type long-range EP coupling
\begin{equation}
f_l(j)=\frac{\kappa}{(|l-j|^2+1)^{3/2}}
\label{krako} 
\end{equation}
(cf. Fig.~1 of Ref.~\cite{AK99}).
The distance $|l-j|$ is measured in units of the lattice constant. 
In (1), $x_0=\sqrt{1/2M\ho}$, $\kappa x_0=\sqrt{\ep\ho}$, and 
the optical phonons, being polarized in the direction
perpendicular to the chain, are approximated as 
independent Einstein oscillators with bare frequency $\ho$ ($\hbar=1$). 
Physically, this model was proposed to mimic the
interaction of doped holes with apical oxygens in 
the high--$T_C$'s, e.g. in $\rm YBa_2Cu_3O_{6+x}$,
where one can assume that the coupling is not screened
because of a low c-axis conductivity and high phonon frequency~\cite{AK99}.
Methodically, model (1) represents an extension of the Fr\"ohlich model to 
a discrete ionic lattice or of the Holstein model including longer ranged
EP interactions. Indeed, defining the polaron binding energy as
\begin{equation}
\eps=\frac{x_0^2}{\ho} \sum_lf^2_l(0)=1.27 \,\ep\;, 
\label{eps}
\end{equation}
the Holstein model (HM) results by setting 
\begin{eqnarray}
f_l(j)&=&\kappa \delta_{j,l}\,,\nonumber\\
\eps&\to& \ep\,.
\label{hmo}
\end{eqnarray}
Therefore, the model (1) will be subsequently  
termed {\it extended Holstein model} (EHM).
In order to parametrize the EP coupling strength for both 
the HM and EHM we introduce two dimensionless EP coupling constants
\begin{eqnarray}
\lambda&=&\eps/2t\,,\nonumber\\
g^2&=&\eps/\ho
\end{eqnarray}
(in what follows we measure all energies in units of $t$).

So far, analytical and numerical investigations of the 
EHM have been mainly confined to the determination
of the effective mass of the polaron, where it was found that 
the EHM polaron is much lighter than the small Holstein
polaron~\cite{AK99}. First results for the polaron band 
dispersion and density of states were obtained quite recently,
however, the QMC method of calculating the ground-state dispersion 
used by Kornilovitch~\cite{Ko99} is limited to the case where the bandwidth
is much smaller than the phonon frequency. 

In this paper we present an detailed comparative study of the 
Holstein and extended Holstein models in order to discuss 
the effects of long-range EP forces on the lattice polaron formation. 
Using exact Lanczos diagonalization supplemented by a well-controlled
phonon Hilbert space truncation method, we  
calculate for the first time spectral (optical) properties of 
the EHM polaron. As stated above such an numerical investigation is especially
valuable in the non--adiabatic intermediate-to-strong coupling  
transition region, where the electronic and phononic energy 
scales are not well separated, i.e., $\lambda \simeq \ho/t \simeq 1$. 
In the weak-- and strong--coupling regimes the numerical work is 
supplemented by analytical approaches outlined in the Appendix. 
\section{Quasiparticle properties of lattice polarons}
\subsection{Numerical Methods}
Before we discuss the various physical quantities let us briefly 
sketch our computational scheme. 
Diagonalizing the coupled EP system 
(1) on finite 1D lattices
with periodic boundary conditions (PBC), a general 
$K$--symmetrized state is 
given as $|{\mit \Psi}_K^{}\rangle = \sum_{m=0}^M
\sum_{\bar{s}=1}^{\bar{S}(m)} c_K^{m,\bar{s}} \,
|K;m,\bar{s} \rangle$, where $\bar{S}(m)=(N-1+m)!/(N-1)!m!$. 
$K$ denotes the total momentum
of the coupled EP system. Because the phonon Hilbert 
space has infinite dimension we apply a truncation procedure 
restricting ourselves to phononic states 
$|m,\bar{s}\rangle_{ph}=\prod_{l=0}^{N-1}
\frac{1}{\sqrt{n_l^{\bar{s}}!}}\left(b_l^\dagger\right)^{n_{l}^{\bar{s}}}\,
|0\rangle_{ph}$
with at most $M$ phonons, whereby $m=\sum_{l=0}^{N-1} n_l^{\bar{s}} \le M$,  
and $n_l^{\bar{s}}\in [0,m]$ (cf. Refs.~\cite{BWF98}).  
The ground state $| {\mit \Psi}_{0,K=0}^{}\rangle$ and all excited states  
$| {\mit \Psi}_{n,K}^{}\rangle$ contain components that correspond to 
$m$--phonon states in the tensorial product Hilbert space 
of electronic and phononic states. 
Accordingly, 
\begin{equation}
|c^m_0|^2(M)=\sum_{\bar{s}}^{\bar{S}(m)}  |c_{K=0}^{m,\bar{s}}|^2
\end{equation}
can be taken as a measure of the weight of the $m$--phonon state 
in the $K=0$ ground state. 
In our ED analysis convergence is assumed to be achieved if the ground--state
energy $E_0(M)$ is determined with a relative error less than $10^{\rm -7}$
and $|c^M_0|^2(M)\leq 10^{\rm -7}$. Afterwards static correlation functions
can be obtained easily by calculating  ground--state expectation values 
$\langle {\mit \Psi}_0^{}(M)|\ldots|{\mit \Psi}_0^{}(M)\rangle$.
The numerical computation of dynamical properties, i.e. of 
spectral functions 
\begin{eqnarray}
A^{\cal O}(\omega)&=&-\lim_{\varepsilon\to 0^+}\frac{1}{\pi} \Im m \left[
\langle{\mit \Psi}_{0}
|{\bf O}^{\dagger}\frac{1}{\omega - {\bf H} +E_0 +i\varepsilon}{\bf O}^{} 
|{\mit \Psi}_{0}\rangle\right]\nonumber\\&=&
\sum_{n=0}^{D-1}|\langle{\mit \Psi}_{n}|{\bf O}^{\dagger}|
{\mit \Psi}_{0}\rangle |^{2}\delta [\omega - (E_{n} - E_{0})]\,,
\end{eqnarray} 
is much more involved. Here ${\bf O}$ denotes the matrix 
representation of a certain operator ${\cal O}$, and  
${\bf H}$ is the very large sparse Hamilton matrix, 
acting in a Hilbert space with fixed momentum, 
which, for our problem, has a typical total dimension 
($D$) of about $10^{8}-10^{9}$.
Since it is impossible to determine all the
eigenvalues $(E_n)$ and eigenstates ($|{\mit \Psi}_{n}\rangle$) 
of such a huge Hamilton matrix we combine kernel 
(Chebyshev) polynomial expansion and maximum entropy optimization
in order to calculate $A^{\cal O}(\omega)$ in a well--controlled
approximation (for more details see Refs.~\cite{BWF98,SR97}). 
\subsection{Electron lattice correlations} 
In a first step we discuss the different nature of the polaronic states
in the HM and EHM in terms of static correlation 
functions $\langle n_iq_l\rangle$ between the 
electron position [$i=0$] and the oscillator displacement [$q_l\propto
(b_l^\dagger+b_l^{})$] at site $l$,
\begin{equation}
\label{elcf}
\chi_{0,l}^{}=\langle c_0^\dagger c_0^{} (b_{0+l}^\dagger 
+ b_{0+l}^{})\rangle/{\cal N}\;.
\end{equation}
$\chi_{0,l}^{}$ indicates the strength of the electron 
induced lattice distortion at $i=0$ and its spatial 
extent~\cite{JW98b,WF98a,BTB99}, where  
${\cal N}=\sum_l \langle c_0^\dagger c_0^{} (b_{0+l}^\dagger 
+ b_{0+l}^{})\rangle$ is a normalization constant 
(note that ${\cal N}=2(\ep/\ho) \langle c_0^\dagger c_0^{}\rangle$
holds for the HM).

Figure ~1 shows the (static) electron--lattice correlation 
function~(\ref{elcf}) in the weak-- (a) and intermediate--to--strong-- (b) 
EP coupling regimes, where we have chosen an intermediate phonon frequency 
($\ho=1$) in order to include non--adiabatic effects.
Clearly for the quantum phonon model~(1) the EP interaction gives 
rise to a ``dressing'' of the charge carrier
at any finite $\lambda$, $g^2$. If the EP coupling is weak, however, 
the amplitude of $\chi_{0,l}^{}$ is small $\forall$ $l$ (in particular 
smaller than the quantum--lattice  (zero--point) fluctuations), 
that means the lattice deformation could not trap the charge carrier 
and  a so--called ``large'' polaron (LP) is formed in both the Holstein 
and extended Holstein models. Obviously, the situation is entirely 
different in the strong--coupling region. For the HM the EP correlations are 
almost local indicating the formation of a ``small'' polaron
(SP). On the other hand, as a result of the non--screened EP interaction, in 
the EHM  the deformation is spread over many lattice sites, i.e.,  
we found again a LP. It is worthwhile to point out, however,  
that the electron and the phonon cloud are tightly bound.  
That means the LP of the EHM as a whole behaves as a well-defined 
polaronic quasiparticle (cf. Sec.~II~C) 
and, in our opinion, it is not possible to discuss 
the size of the electronic wave function
and the size of the lattice distortion separately~\cite{AK99}.     

In the insets of Fig.~1 we show the differences 
between the phonon distribution functions in the weak-- 
and strong--coupling cases, where the ground state 
is basically a zero--phonon-- and multi--phonon state,
respectively. With regard to the discussion of the effective
mass in Sec.~II~E we would like to annotate here, 
that at small (large) $\lambda$ the EHM polaron contains 
more (less) phonons in its phonon cloud than the HM polaron.  
Of course, in the extreme strong-coupling limit, 
the usual Poisson distribution with parameter $g^2$ results,
demonstrating that adjusting the parameters of both models
according to~(3)--(5) is correct.   
\subsection{Single--particle spectral function}
Next, in order to examine dynamical quasiparticle--properties of
the HM and EHM polarons, we have evaluated the wave--vector 
resolved spectral density function  
\begin{equation}
  A_{K}(E) = \sum_n |\langle {\mit\Psi}_{n,K}^{} 
\,|\,c_{K}^{\dagger}\,|\,0\rangle|^2\,\delta ( E-E_{n,K})\,.
\end{equation}
The results are presented in Figure 2. To visualize the 
spectral weights of the various excitations,   
the integrated density of states,
\begin{equation}
 N_{K}(E)=\int_{-\infty}^{E} dE' A_{K}(E')\,,
\end{equation}
is also displayed. The weight of the first delta--function 
peak in each $K$--sector gives the wave--function 
renormalization factor~\cite{Mah90}
\begin{equation}
Z_K= |\langle {\mit\Psi}_{0,K}^{} 
\,|\,c_{K}^{\dagger}\,|\,0\rangle|^2\;,
\end{equation}
where $|{\mit \Psi}_{0,K}\rangle$ denotes the single--polaron 
state with momentum $K$ being lowest in energy.
$Z_{K=0}$ is usually termed ``quasiparticle--weight factor''.   
Since the total integrated area under the entire spectra is unity,
the renormalization factor is less than unity and, in particular, 
$Z_{K=0}$ is a measure how much the polaronic quasiparticle
``deviates'' from the free electron ($Z_{K=0}=1$).
In accordance with the discussion in the preceding section, 
in the weakly--interacting EHM we found a stronger dressing of 
the electron by phonons than in the HM, i.e., a larger renormalization 
$Z_{K=0}^{EHM}<Z_{K=0}^{HM}$. 
This finding is corroborated  by the the weak--coupling theory (WCT)
outlined in Appendix A.  Table~I demonstrates the good 
agreement of the theoretical approach, working for the infinite system,
and finite--cluster diagonalizations, provided that both $\lambda$ and
$g^2$ are small. Contrary to $Z_{K=0}$, which is only slightly reduced  
from the free electron value, the wave--function renormalization factor
$Z_{K=\pi}$ is almost zero. The WCT shows that the state with $K=\pi$,
being  energetically separated by $\ho$ from 
the ground--state energy, is predominantly a phononic state.
At strong EP coupling the polaronic band is characterized by 
$Z_{K}\ll 1$ $\forall K$, indicating a strong mixing of electronic 
and phononic degrees of freedom. Calculating the polaronic
quasiparticle weight factor within the framework 
of the strong--coupling theory (SCT) developed in Appendix B 
(Eq.~(\ref{swfsc})) gives $Z_{K=0}$--values 
which are by a factor of 3 too small 
as compared to the exact data of Fig.~2 (b) and (d). The differences
mainly arise because these parameters correspond rather to the
intermediate-to-strong than to the extreme strong--coupling regime. 
The qualitative behavior
of the single--particle spectral function $A_K(E)$, however, is 
correctly reproduced by SCT~(\ref{specfunsc}), which yields, 
above the quasiparticle pole, a sequence of excitations
separated by $\ho$. Apparently the spectral weight of this incoherent
part increases with increasing EP interaction strength.
\subsection{Polaron band structure}
Now the so--called ``coherent'' band dispersion, $E_K$, can be derived 
from the first peak of $A_{K}(E)$. Figure~3 compares the 
dispersion of the energy bands for the EHM at different EP interactions, 
corresponding to the weak--, intermediate-- and strong--coupling case,
where, by going from (a) to (c), $\ho$ is increased modelling 
the adiabatic, intermediate, and anti-adiabatic regimes. 

Starting with the adiabatic weak--coupling case (Fig.~3~a), we found that 
the band structure is nearly unaffected at small momentum, i.e.
in the vicinity of the band center. In this region 
$|\Psi_{0,K}^{}\rangle$ is quasi a zero--phonon state~\cite{WF97,Ro98}. 
A different behavior is observed near the zone boundary. 
Here the band structure is flattened. 
Such a ``flattening'' has been found for the HM as well    
and can be attributed to the intersection of 
the dispersionsless optical phonon branch with 
the bare electronic cosine band~\cite{LR73,St96,WF97,Ro98}
(cf. also Fig.~4). The weak--coupling calculation of $E_K$ 
(Appendix ~A; Eq.~(\ref{bdwc})) reflects this behavior.   
For the HM, e.g., the correction to $\xi_K$ is given by the
integral ${\cal I}^{(1)}(K,0)$, which is non--zero
only for $(E_K-\ho)^2> 4t^2$. The latter condition yields a threshold $K^*$,
at which the solution of~(\ref{bdwc}) jumps to the bare band dispersion 
$\xi_K$. For $K>K^*$, the first excitation in $A_K(E)$ 
is related to a one--phonon absorption process and   
as a result the WCT approximation for $E_K$ breaks down. 
Thus, above $K^*$, the physical solution is given to 
lowest order by the dashed line at $E_0+\ho$.
 
The flattening considerably weakens and ultimately vanishes 
if the EP coupling $\lambda$ increases. This tendency is especially 
pronounced for the EHM in the non--adiabatic regime.       
As can be seen from Figs.~3 (b) and (c), in the strong--coupling 
non--adiabatic regime our SCT yields excellent results. 
Most notably we do not observe the same drastic polaronic 
band collapse as in the HM~\cite{WF97,FLW97}, i.e., in the EHM  
the coherent bandwidth $\Delta E=E_{\pi} - E_0$ becomes much 
less renormalized by the EP interaction (e.g., for the HM with
$\lambda=5.0$ and $\ho =3.0$, we found $\Delta E =0.15319$).
\subsection{Effective mass}
Another important question is the change in the polaron effective mass induced
by the EP coupling. In general it is difficult to compute the mass enhancement,
which is defined as an inverse second derivative of the band energy
with respect to quasi--momentum at the band minimum,
$m^*/m\propto [\partial^2E(K)/\partial K^2|_{K=0}]^{-1}$,
using finite--lattice diagonalizations, because $E(K)$ is
known at multiples of $2\pi/N$ only rather than at any $K$, 
making the limiting procedure $K\to 0$ ill posed.
On the other hand, the mass enhancement factor $m^*/m$ is also 
related to the quasiparticle weight factor $Z_{K=0}$~\cite{Mah90}. 
For the HM, we are able to  prove the relation 
\begin{equation}
\label{meffhomo}
m/m^*_{HM}=Z_{K=0}^{HM}
\end{equation}
in the weak--coupling limit (see Appendix~A, 
Eqs.~(\ref{zmahan2})--(\ref{meffwc})), i.e., at $\lambda\ll 1$     
the polaron effective mass can be read off from 
the first step in the integrated spectral weight function 
depicted in Fig.~2. Plotting $Z_{K=0}^{HM}$ as a function
of $\lambda$  in Fig.~5 and comparing the effective mass determined
in this way with the QMC masses, obtained from 
$m/m^*=\partial^2E_K/\partial K^2$ without 
any systematic finite--size errors~\cite{KP97,AK99,Ko99}, 
the perhaps surprise finding is that Eq.~(\ref{meffhomo}) 
holds for the {\it whole} coupling region. 
That means, in the Holstein model, we can determine the effective
polaron mass simply by calculating the quasiparticle weight factor. 
In previous ED studies of the Holstein polaron problem this fact 
has been ignored so far.

Unfortunately no such simple relation exists for the EHM. 
This is shown more explicitly in the Appendix, 
where approximative expressions for $m/m^*$ 
were derived in the weak-- and strong--coupling limits. Note that
our analytical weak-- and strong--coupling approaches confirm the
unexpected non-monotonic  dependence on $\lambda$ of 
$m^*_{EHM}/m^*_{HM}$, which was found numerically  by Alexandrov and 
Kornilovitch~\cite{AK99} (see inset).  In the light of the results 
presented in the previous sections, it becomes clear that at small
EP couplings the EHM LP has to drag a larger phonon cloud coherently
through the lattice than the HM LP and therefore acquires a larger
effective mass. Further numerical data show that 
this effect becomes negligible in the weak--coupling
anti--adiabatic regime, where the phonons can follow the
electron instantaneously (cf. also Fig.~3 of Ref.~\cite{AK99}). 
As a matter of course, in the strong--coupling limit
the EHM LP is much lighter than the HM SP due to the 
weaker band renormalization caused by the extended 
form of the lattice distortion.  
\subsection{Optical conductivity}
In this section we compare the optical response of HM and EHM polarons.
The real part of the optical conductivity, 
\begin{equation}
\label{resigma}
\mbox{Re}\sigma(\omega)=\cD\delta(\o)+ \sigma^{reg}(\omega)\;,
\end{equation}
can be decomposed into the Drude term ($\propto \cD$) at $\o=0$ and a regular
contribution for $\o>0$, which,  in linear response theory, 
for the (extended) Holstein model is given by
\begin{eqnarray} 
\sigma^{reg}(\omega)&=&\frac{\sigma_0}{N}\sum_{n \neq 0}
\frac{|\langle {\mit \Psi}_{0}^{} |i t\sum_{j}( c_{j}^{\dagger}
 c_{j+1}^{} - c_{j+1}^{\dagger}c_{j}^{}) |  {\mit \Psi}_{n}^{} 
       \rangle |^2}{E_n-E_0}\nonumber\\
 &&\hspace*{2cm}\times\;\delta[\omega -(E_n-E_0)] 
\end{eqnarray}
with $\sigma_0=\pi e^2$ ($T=0$, $K=0$--sector). 
Again we introduce a $\o$-integrated spectral weight function, 
\begin{equation}
\cS^{reg}(\o)=\int_0^{\o} d\o' \sigma^{reg}(\o')\;,
\end{equation}
in order to visualize the intensity of the various excitations 
more clearly. 

Figure~6 shows the optical conductivity obtained at $\ho=1.0$ 
for the HM and EHM on a eight--site lattice using PBC. 
For the 1D HM the optical absorption spectrum has been discussed in 
detail in previous work~\cite{WF98a}. If the energy to excite one phonon
lies inside the bare tight--binding band we found, at weak EP coupling,
the first transitions by adding phonons with opposite momentum 
to only slightly renormalized electronic states (in order to reach the
$K=0$ sector of the ground state). However, since the ground state 
is approximately a zero--phonon state (cf. inset of Fig.~1 ~(a)), 
the spectral weight of optical transitions involving larger
number of phonons is reduced drastically.  Of course, the absorption 
threshold is $\ho$ for the infinite system; the shift observed in Fig.~6~(a) 
simply results from the discrete $K$--mesh of our finite system.    
The situation changes by increasing the EP coupling when 
in the HM the SP formation takes place. Now the phonon
distribution function in the ground state is broadened and 
in the optical response the overlap with excited multi--phonon states 
is enlarged. As a result the famous SP absorption maximum 
develops around $\omega\simeq4\lambda=2\eps$ for large couplings.

Let us now discuss the optical response in the EHM. 
Of course, there is little change in the weak--coupling region.
At large EP coupling, however, the optical absorption   
points toward a completely different nature of the polaronic states in the 
Holstein and extended Holstein models. The EHM polaron clearly shows
all the LP signatures but compared to the weak--coupling case and,
what is more important, also compared to the HM SP,   
the optical conductivity is strongly enhanced 
due to by multi--phonon absorptions processes.   
The physical reason lies in the non--screened EP
interaction leading to the form of the lattice distortion 
depicted in Fig~1~(b). Taking into account 
the internal structure of the EHM LP it is obvious 
that the lattice distortion undergoes less relative changes 
when the charge carrier hops incoherently to neighboring sites  
accompanied by phonon absorption or emission~\cite{AK99}.  
\subsection{Kinetic energy}
Integrating~(\ref{resigma}) with respect to  $\o$, the familiar 
\mbox{f-sum} rule
\begin{equation}
\label{suru}
- \frac{E_{kin}}{2}=\frac{\cS^{tot}}{\sigma_0}=\frac{\cD}{2\sigma_0}
+\frac{\cS^{reg}}{\sigma_0}\;,
\end{equation}
can be derived, where $\cS^{reg}=\cS^{reg}(\infty)$. 
Eq.~(\ref{suru}) relates optical response and kinetic energy.
$E_{kin}$ measures the mobility of the charge carrier.

To elucidate the different nature of HM and EHM polarons in more detail, 
in Fig.~7 we have displayed the kinetic energy ($\propto \cS^{tot}$), 
renormalized to its value at $\lambda=0$, together with $\cS^{reg}$.
Since the kinetic energy contains contributions from both 
``coherent'' ($\propto \cD$)  and ``incoherent'' 
($\propto \cS^{reg}$) hopping processes,  the Drude part can 
be directly read off from the difference between the filled and 
open symbols at fixed $\lambda$. In agreement with previous numerical 
results, in the HM we found a continuous transition from a LP to a 
less mobile SP as the EP interacting increases~\cite{RL82,WF98a,JW98b}. 
The decrease of $S^{tot}$ in the crossover region, being much 
more pronounced in the adiabatic regime~\cite{WF98a} (as well as  
for higher dimensions~\cite{RL82,FLW97}), is driven by the sharp 
drop of the Drude weight~\cite{CSG97}. 
By contrast, in the EHM the kinetic energy decreases very gradually 
with increasing $\lambda$ and we observe a substantial Drude contribution
even at large EP couplings. This is in accord with the moderate
renormalization of the polaronic bandwidth and the minor effective mass
(cf. Sec.~II~D and Sec.~E). In addition, as already stressed in Sec.~II~E, 
the optical absorption due to inelastic scattering processes, 
described by the regular part of the optical conductivity, 
gives a large contribution. This can be easily understood within 
second--order perturbation theory (note that one has to go beyond
the lowest order of approximation to obtain reliable results for 
the kinetic energy~\cite{FK97,St96,FLW97}; cf. Appendix~B): during 
a second--order hopping process of the EHM LP the lattice distortion 
of the intermediate state with the charge carrier on a nearest--neighbor     
site of the initial site fits much better to the polaronic quasiparticle  
than in the case of the HM SP.  
The difference between the numerical and theoretical results
at larger EP couplings originates from the neglect of longer--ranged
hopping processes in our theoretical approach. Of course, such 
transitions are much more important in the EHM.    
In addition, let us emphasize that $S^{tot}_{EHM}>S^{tot}_{HM}$ 
holds also in the weak--coupling limit (see inset).
That means the stronger reduction of the coherent Drude part, 
corresponding to the stronger mass enhancement in the weakly--coupled
EHM ($m^*_{EHM}/m^*_{HM}>1$,  cf. inset Fig.~5), 
is overcompensated by the incoherent part.   
\section{Conclusions}
In summary, we have performed an extensive comparative numerical
study of polaron formation in the Holstein and extended Holstein 
models, supplemented by a theoretical analysis of 
the weak-- and strong coupling limits. 
The emphasis was on the new effects induced by a non--screened
electron--phonon interaction. The main characteristics of the
new polaronic state formed in the EHM are the following.
\begin{itemize}
\item[(i)] By its nature the EHM polaron is a large polaron in the whole EP 
coupling region. That is the lattice distortion is spread 
over large distances even if the EP is extremely strong, 
In this regime a small polaron is formed in the Holstein model.
\item[(ii)] For strong EP interactions the EHM polaron propagates
in a relatively weakly renormalized band as compared to the HM. 
Accordingly the effective mass of the large EHM polaron is much 
smaller than that of the small Holstein polaron  with 
the same polaron binding energy.     
\item[(iii)] A surprise finding is that the effective mass of the
EHM polaron, describing a ``coherent'' band motion, is larger than the 
effective mass of the HM polaron at weak EP couplings, 
in particular in the adiabatic case.  
We have seen that this effect can be attributed to the larger
number of phonons the charge carrier has to drag through 
the lattice if the weak EP interaction is non--screened.
\item[(iv)] From the calculation of the $K$--resolved
single particle spectral function a wave--vector renormalization 
factor $Z_K$ can be extracted, which indicates, 
in accordance with (i) and (ii), at weak (strong) EP couplings  
a stronger (weaker) renormalization of the band states 
in the EHM than in the HM.  
\item[(v)] While in the HM the inverse polaron effective mass 
is directly given by the quasiparticle weight factor, $Z_{K=0}$,
the relation is more complicated for the EHM. In the weak--coupling 
limit this has been corroborated analytically.    
\item[(vi)]  
The EHM polaron band dispersion is non--cosine for model parameters 
corresponding to the (adiabatic) weak--to--intermediate coupling regime.  
In particular, in the weakly--coupled EHM, a flattening of the band structure 
at the zone boundary is definitely observed just as in the HM, 
but the effect is much less pronounced. Furthermore the 
flattening rapidly vanishes with increasing EP coupling 
strength and phonon frequency. 
In the strong--coupling limit, the EHM exhibits a free--particle--like 
band dispersion with a bandwidth which, although renormalized, is 
approximately one or two orders of magnitudes larger than in the HM.  
\item[(vii)] While in the HM the transition from large to small polarons 
is accompanied by significant changes in the optical response, the 
optical absorption in the EHM shows large polaron characteristics
for all EP interaction strengths. Most notably the extended form of the 
the lattice distortion in the EHM gives rise to a large amount
of ``incoherent'' hopping  processes contributing to 
the regular part of the optical conductivity. As a result the regular 
contribution to the f--sum rule is always bigger than in the HM.
\item[(viii)] If one takes the averaged kinetic energy as a measure
for the mobility of a charge carrier, the EHM polaron is more 
mobile than the HM polaron, independently of the magnitude of the
EP coupling strength. In particular the dramatic kinetic energy loss 
during the self--trapping transition of the Holstein small polaron 
is absent in the EHM. On the contrary, one observes a very gradual
decrease of the kinetic energy with increasing EP interaction and
a substantial Drude contribution even for large EP couplings.
\end{itemize}
Finally we would like to stress that the above properties of
the EHM large polaron are generic and not an artifact of our  
1D system. The relative mass enhancement
$m^*_{EHM}/m^*_{HM}$ at weak EP interactions, e.g., 
seems to be even more pronounced in 2D~\cite{AK99}.
\section*{Acknowledgements}
The authors are greatly indebted to P. Kornilovitch
for putting his QMC data at our disposal. 
Numerical calculations were performed at the LRZ M\"unchen, 
NIC J\"ulich, and the HLR Stuttgart.
This work was supported by the Deutsche Forschungsgemeinschaft
and the Czech Academy of Sciences under Grant No. 436 TSE 113/33.
\section*{Appendix: Analytical approaches}
Following the previous consideration of the HM~\cite{FLW97}, the 
Hamiltonian for treating the EHM at any $\lambda$ may be written as
\begin{equation}
\cH = - \eta\sum_j c^\dagger_j c^{}_j 
-\sum_{j,j'}{\cal C}_{j^\prime j} c_{j'}^{\dagger} c_{j}^{}
+\ho\sum_l  ( b_l^{\dagger} b_l^{}+ 
\mbox{\small $\frac{1}{2}$})\,,
\label{EDA}
\end{equation}
where $\eta$ is a c--number and $\cC_{j^\prime j}$ are generally functions
of the phonon operators $b_l^{\dagger}$, $b_l^{}$. Using the formalism of 
generalized Matsubara Green's functions, the polaron self-energy $\mS$
corresponding to~(\ref{EDA}) in the second step of iteration is given 
by 
\begin{eqnarray}
\label{SEA}
\mS(j_1 \tau_1;j_2\tau_2)&=&-\langle\cC_{j_1j_2}\rangle 
\delta(\tau_1-\tau_2)+\sum_{j^\prime j^{\prime\prime}}
\cG(j^\prime\tau_1;j^{\prime\prime}\tau_2)\times
\nonumber\\
&&\hspace{-1cm}\times[\langle\cT_{\tau}\cC_{j_1j^\prime}
(\tau_1)\cC_{j^{\prime\prime}j_2}(\tau_2)
\rangle -\langle\cC_{j_1j^\prime}\rangle
\langle\cC_{j^{\prime\prime}j_2}\rangle]
\,,
\end{eqnarray}
where $\cG(j^\prime\tau_1;j^{\prime\prime}\tau_2)$ means the polaron
Green's function in the first approximation
(see Refs.~\cite{Sc66,Lo94,KB62} for details). 
In the subsequent calculations, the latter equation will be 
taken as a starting point for the treatment of EHM in the 
weak $(\lambda\ll 1)$-- and strong
$(\lambda\gg 1)$--coupling regimes; in addition, the low--temperature
approximation  (LTA) defined by $\beta\ho\gg 1$, and the small 
carrier--concentration limit $(x\to 0)$ will be assumed.  
\subsection{Weak--coupling regime}
In this case, the functions $\cC_{j^\prime j}$ are defined by the 
first and third terms of the Hamiltonian~(\ref{ED1}), and $\eta$
is put equal to the chemical potential $\mu$. Applying the Fourier
transformation to both sides of~(\ref{SEA}) and carrying out the 
standard summation over the phonon Matsubara 
frequencies~\cite{Mah90,FLW97}, the polaron self--energy
\begin{eqnarray}
\label{seenwc}
\mS_K(\bar{\o})&=&\xi_K+\ho\sum_{d=-\infty}^\infty \eps(d) \cos (Kd)\\
&&\hspace*{.8cm}\times\;
\int_{-\pi}^{\pi}\frac{dK'}{2\pi}
\frac{\cos (K'd)}{\bar{\o}+\mu-\ho-\xi_{K'}}\nonumber
\end{eqnarray}
is obtained after analytical continuation to the real
frequencies $\omega$ under assumption of LTA and small $x$.
In writing~(\ref{seenwc}), the definition
\begin{equation}
\label{epsd}
\eps(d)=\eps\sum_l f_l(0)f_l(d)/\sum_l f_l^2(0)\,,
\end{equation}
$\xi_K=-2t\cos K$, and $\bar{\o}=\omega+i0^+$ were used. The 
HM limit results from~(\ref{seenwc}) by setting
\begin{equation}
\label{holi}
\eps(d)=\ep \delta_{d,0}^{}\,.
\end{equation}

In view of~(\ref{seenwc}), the {\it polaron band energies} 
$E_K$ are solutions
of the following equation 
\begin{equation}
\label{bdwc}
E_K=\xi_K+\ho\sum_{d=-\infty}^\infty \eps(d)\, \cos (Kd)\, 
{\cal I}^{(1)}(K,d)\,,
\end{equation}
and the {\it renormalization factor of the spectral function}~\cite{Mah90},  
$Z_K=\left[1-\frac{\partial}{\partial \omega} \mbox {Re} 
\mS_K(\o)\right]^{-1}_{\omega=E_K-\mu}$, is determined by 
\begin{equation}
\label{zmahan2}
Z^{-1}_K=
1+\ho\sum_{d=-\infty}^\infty \eps(d) \,\cos (Kd)\,{\cal I}^{(2)}(K,d)\;,
\end{equation}
where
\begin{equation}
\label{in}
{\cal I}^{(n)}(K,d)=\int_{-\pi}^{\pi}\frac{dK'}{2\pi}
\frac{\cos (K'd)}{(E_K-\ho-\xi_{K'})^n}\,.
\end{equation}

The relation between the {\it effective polaron mass} $m^*$
and the bare electron mass $m$ (being equal to $(2t)^{-1}$ in 1D)
is deduced according to $m/m^*=[\partial E_K/\partial 
\varepsilon_K]|_{\varepsilon_K\to 0}$ 
with $\varepsilon_K= t K^2$~\cite{Mah90}:
\begin{equation}
\label{meffwc}
\frac{m}{m^*}=Z_{K=0}\left[1-
\frac{\ho}{2t}\sum_{d=-\infty}^\infty 
\eps(d)\,d^2\,{\cal I}^{(1)}(0,d)\right]\,.
\end{equation}
Note that only for the Holstein model $(d=0)$ the effective mass is given 
by the inverse spectral weight factor 
$m^*_{\rm HM}/m=\left(Z_{K=0}^{\rm HM}\right)^{-1}$.
Using~(\ref{bdwc}), the {\it polaron kinetic energy} results 
from the relation 
$E_{kin}=t\partial_tE_K|_{K\to 0}$ as 
\begin{eqnarray}
\label{kinenwc}
E_{kin}&=&- 2tZ_{K=0}\left[1+\ho\sum_{d=-\infty}^\infty 
\eps(d)\right.\\
&&\hspace*{1.5cm}\left.\times
\int_{-\pi}^{\pi}\frac{dK'}{2\pi}
\frac{\cos K'\cos (K'd)}{(E_0-\ho-\xi_{K'})^2}\right]\,.\nonumber  
\end{eqnarray}
\subsection{Strong--coupling regime}
In order to generalize the strong-coupling approach, developed in~\cite{FLW97} 
for the HM, to the EHM case, the long-range interaction of 
Hamiltonian~(\ref{ED1}) is eliminated by a non--local 
Lang--Firsov transformation 
\begin{equation}
\cU=\prod_l\exp\left\{\frac{x_0}{\ho}\sum_{j} 
f_l(j)c_{j}^{\dagger} c_{j}^{}
( b_l^{\dagger}  - b_l^{})\right\}\,.
\label{utr}
\end{equation}
Clearly, the theory based on~(\ref{utr}) turns into the theory for the  HM 
if the condition~(\ref{hmo}) is inserted. 
The canonical transformation~(\ref{utr})
applied to~(\ref{ED1}) leads to the polaron binding energy~(\ref{eps}) and 
to the emergence of the multi--phonon processes connected with the electron 
hopping from the site $j$ to the nearest-neighbor sites  $j+h$ ($h$ being the 
elementary translation in units of the lattice constant).

Treating the dynamical EP interaction of the transformed Hamiltonian 
by means of the formalism outlined in~\cite{FLW97}, the polaron 
self--energy represented in the space of Brillouin--zone $K$--vectors
and Matsubara frequencies $i\o_\nu =i(2\nu+1)\pi/\beta$ is obtained 
to the second order as
\begin{eqnarray}\label{L6}
\mS_K(i\o_\nu)&=&-t\sum_{h}\langle\mP_{j,j+h}\rangle
\mbox{e}^{iKh}+\sum_{j',j''\atop j_2-j_1}\mbox{e}^{iK(j_2-j_1)}\nonumber\\
&&\hspace*{-1cm}\times\!
\frac{1}{N}\sum_{K',\zeta}\mbox{e}^{iK'(j'-j'')}
\frac{1}{\beta}\int_0^\beta d\tau\mbox{e}^{i(\o_\nu-\o_\zeta)\tau}
\\
&&\hspace*{-2cm}
\times \;t^2\Big(\langle\mP_{j_1,j'}(\tau)\mP_{j'',j_2}(0)\rangle
-\langle\mP_{j_1,j'}\rangle\langle\mP_{j'',j_2}
\rangle\Big)\cG_{K'}(i\o_\zeta)\;.
\nonumber
\end{eqnarray}
The multi--phonon operator
\begin{equation}
\mP_{j,j+h}=\exp\left\{\frac{x_0}{\ho}\sum_{l} 
\Big(f_l(j+h)-f_l(j)\Big)\Big( b_l^{\dagger}  - b_l^{}\Big)\right\}
\label{phi}
\end{equation}
occurring in the transformed hopping term of~(\ref{EDA}) induces 
the polaron band narrowing in the first order, namely,  
\begin{equation}
\ts\equiv t\langle\mP_{j,j+h}\rangle=t\exp\{-g^2\Delta(1)
\coth (\beta\ho/2)\}\;,
\label{ts}
\end{equation} 
where
\begin{equation}
\Delta(1)=\left[
1-\sum_l f_l(0) f_l(1)/ \sum_l f_l^2(0)\right]\;.
\label{gs}
\end{equation} 
In the LTA $\beta\ho\gg 1$, we have   
$\ts\simeq t\exp\{-g^2\Delta(1)\}$, $\tilde{\xi}_K=-2\ts\cos K$, and 
\begin{eqnarray}
\mS_K(i\o_\nu)&=&\tilde{\xi}_K
+\;\sum_{j',j''\atop j_2-j_1}\mbox{e}^{iK(j_2-j_1)}
\frac{1}{N}\sum_{K',\zeta}\mbox{e}^{iK'(j'-j'')}
\nonumber\\
&&\hspace*{-1cm}\times\;
\tsq\sum_{s\geq 1}\frac{\vk^s}{s!}\frac{1}{\beta}
\frac{2s\ho}{(s\ho)^2+(\o_\zeta-\o_\nu)^2}\cG_{K'}(i\o_\zeta)\,.
\label{seen2}
\end{eqnarray}
The parameter $\vk\equiv\vk(d,d_1,d_2)$ is given by
\begin{eqnarray}
\vk&&=\frac{x_0^2}{\ho^2}
\sum_l f_l(0)\Big[
f_l(d)+f_l(d+d_2-d_1)\nonumber
\\&&\hspace*{1.5cm}-f_l(d-d_1)-f_l(d+d_2)\Big]\;.
\label{vk}
\end{eqnarray}
Here $d=j_2-j_1$, $d_1=j'-j_1$ and $d_2=j''-j_2$ are 
elementary translations. 
The summation over $\zeta$ is now evaluated using 
the lowest--order approximation for the Green's function
$ \cG_{K}(i\o_\zeta)$
corresponding to a quasiparticle energy
spectrum of the form  $\tilde{\xi}_K-\eps$.
After carrying out this summation in LTA and for small carrier
concentration, the wave--vector-- and frequency--dependent
polaron self--energy is obtained from~(\ref{seen2}) by the analytical  
continuation $i\o_\nu\to \bar{\o}$ as
\begin{eqnarray}\label{seen3}
\mS_K(\bar{\o})&=&\tilde{\xi}_K
+\tsq\!\!\sum_{d,d_1,d_2}\mbox{e}^{iKd}\frac{1}{N}
\sum_{K'}\\
&&\hspace*{-.5cm}\times\;
\sum_{s\geq 1}\frac{\vk^s}{s!}\mbox{e}^{-iK'(d+d_2-d_1)}\frac{1}{\bar{\o}
-\tilde{\xi}_{K'}+\eps+\mu-s\ho}\;.\nonumber
\end{eqnarray}
If we neglect $\tilde{\xi}_{K^\prime}$ on the right hand side of~(\ref{seen3}),
in the strong--coupling limit, the {\it polaron band dispersion}~$E_K$ can be 
easily determined from 
\begin{eqnarray}
\label{banddissc}
E_K&=& \tilde{\xi}_K -
2\tsq \sum_{s\geq 1}\frac{(2g^2\Delta(1))^s}{s!}
\frac{1}{s\ho-E_K}\\
&&+\ts \tilde{\xi}_{2K}
\sum_{s\geq 1}\frac{[g^2\Delta(2)]^s}{s!}
\frac{1}{s\ho-E_K}\;,\nonumber 
\end{eqnarray}
where
\begin{equation}
\Delta(2)=\left[1
-\frac{2\sum_lf_l(0)f_l(1)-\sum_lf_l(0)f_l(2)}
{\sum_lf_l^2(0)}\right]\,.
\end{equation} 

On the basis of~(\ref{banddissc}), the {\it polaron mass enhancement} 
can be calculated from
\begin{equation}
\label{meffcasc} 
m/m^*=\mbox{e}^{-g^2\Delta(1)}[\partial E_K/\partial 
\tilde{\varepsilon}_K]|_{\tilde{\varepsilon}_K\to 0}\,,
\end{equation} 
where $\tilde{\varepsilon}_K=\ts K^2$.
Substituting, for $K\to 0$, $\tilde{\xi}_K=-2\ts+\tilde{\varepsilon}_K$ 
and $\ts\tilde{\xi}_{2K}=-2\ts(\ts-2\tilde{\varepsilon}_K)$,  
we find
\begin{eqnarray}\label{meff}
\frac{m}{m^*}&=& \mbox{e}^{-g^2\Delta(1)} {\cal Z}^{-1}_{E_0}\\[0.1cm]
&&\hspace*{.5cm}\times\left[1+4t\mbox{e}^{g^2(\Delta(2)-\Delta(1))} 
\left\langle\frac{1}{s\ho-E_0}\right\rangle_{g^2\Delta(2)}\right]\nonumber
\end{eqnarray}
with
\begin{eqnarray}
\label{zzeta}
{\cal Z}_{E_K}&=&\left[1+2t^2\left(
\left\langle\frac{1}{(s\ho-E_K)^2}\right\rangle_{2g^2\Delta(1)}
\right.\right.\\&&\left.\left.
+\mbox{e}^{g^2(\Delta(2)-2\Delta(1))} 
\left\langle\frac{1}{(s\ho-E_K)^2}\right\rangle_{g^2\Delta(2)}\cos 2K\right)
\right].\nonumber
\end{eqnarray}
Here $\langle \ldots \rangle_{\varrho}$ denotes the average over
$s\geq 1$ with respect to the Poisson distribution 
with the parameter $\varrho$. 

Accordingly, the {\it polaron kinetic 
energy} takes the form
\begin{eqnarray} 
E_{kin}&=& -2 t {\cal Z}^{-1}_{E_0}
\left[\mbox{e}^{-g^2\Delta(1)} 
+ 2t \left\langle\frac{1}{s\ho-E_0}\right\rangle_{2g^2\Delta(1)}\right.\\
&&\hspace*{1.4cm}\left.+2t \mbox{e}^{g^2(\Delta(2)-2\Delta(1))} 
\left\langle\frac{1}{s\ho-E_0}\right\rangle_{g^2\Delta(2)}\right]\;.\nonumber
\end{eqnarray}

Finally, in order to discuss qualitatively the behavior of 
the single--particle spectral function reported in~Sec.~II~C, 
we calculate $A_K(\omega)$ in the strong--coupling limit. 
Applying the transformation~(\ref{utr}), the 
electron operators are transformed into 
\begin{equation}
\tilde{c}_j^{(\dagger)}=\exp\left\{(-)\frac{x_0}{\ho}\sum_{l} 
f_l(j)( b_l^{\dagger}  - b_l^{})\right\}c_j^{}\,.
\label{utrc}
\end{equation}
Consequently the spectral function is determined by the imaginary part
of the retarded Green's function $\tilde{\cal G}^R(K,\omega)$ of the 
operators $\tilde{c}_K^{}$, $\tilde{c}_K^{\dagger}$. Owing to the relation
between the time--ordered products of operators and the retarded Green's
functions of the same operators~\cite{DS74} we consider 
the time ordered product 
$\langle{\cal T}_t\tilde{c}_K(t)\tilde{c}_K^{\dagger}(0)\rangle$ 
and perform a decoupled average over the phonon-- and charge
variables. In the limit $T\to 0$ we get 
\begin{eqnarray}
\label{togf}
\langle{\cal T}_t\tilde{c}_K(t)\tilde{c}_K^{\dagger}(0)\rangle&=&
\mbox{e}^{-g^2}\Big[\langle{\cal T}_t c_K(t)c_K^{\dagger}(0)\rangle+
\frac{1}{N}\sum_{K'}\sum_d\sum_{s\geq 1} \nonumber\\
&&\hspace{-2.3cm}
\times\frac{1}{s!}\left(\frac{\eps(d)}{\ho}\right)^s
\mbox{e}^{-is\ho t}\mbox{e}^{i(K-K')d}
\langle{\cal T}_t c_{K'}(t)c_{K'}^{\dagger}(0)\rangle\Big]\,.
\end{eqnarray}
The relation between the imaginary parts of retarded Green's functions
$\tilde{\cal G}^R(K,\omega)$ and ${\cal G}^R(K,\omega)$ of operators
$\tilde{c}_K$ and $c_K$, respectively, is obtained from the 
Fourier transformation of Eq.~(\ref{togf}). As a result,
the spectral function 
\begin{eqnarray}
\label{specfunsc}
\tilde{A}_K(\omega)&=&-2\mbox{Im} \tilde{\cal G}^R_K(\omega)\\
&=&\mbox{e}^{-g^2}{\cal Z}_{E_K}^{-1}2\pi
\delta(\omega-[E_K-\ep-\mu])\nonumber\\
&&\hspace*{-.5cm}+\mbox{e}^{-g^2}
\sum_d\sum_{s\geq 1}\frac{1}{N}\sum_{K'}\frac{1}{s!}
\left(\frac{\eps(d)}{\ho}\right)^s\cos[(K-K')d]\nonumber\\
&&\hspace*{1.3cm}
\,\times {\cal Z}_{E_{K'}}^{-1} 2\pi
\delta(\omega-[E_{K'}+s\ho-\ep-\mu])\nonumber
\end{eqnarray}
is determined by the spectral functions $A_{K'}(\omega-r\ho)$ 
with $r\geq 0$ corresponding to ${\cal G}_{K'}^R(\omega-r\ho)$, i.e.,
using the self-energy $\Sigma_{K'}(\omega')$ derived in this section.
The energies $E_{K'}$ on the r.h.s. of~(\ref{specfunsc}) are solutions 
of~(\ref{banddissc}) and the factors ${\cal Z}_{E_{K'}}$ are given
by~(\ref{zzeta}). The first term on the r.h.s. of~(\ref{specfunsc})
describes the quasiparticle of momentum $K$ and the second term,
being a sum over the entire polaron band, corresponds to the incoherent
part of the spectral function. Inserting the condition~(\ref{holi})
into~(\ref{specfunsc}), the spectral function of the Holstein model
is obtained, which differs by the self--consistently determined
$E_{K'}$ and ${\cal Z}_{E_{K'}}$ from the result 
of Alexandrov and Ranninger~\cite{AR92a}. 

In particular, at $K=0$, $r=0$,  
the {\it quasiparticle weight factor} results as   
\begin{equation}
\label{swfsc}
Z_{K=0}=\mbox{e}^{-g^2}{\cal Z}_{E_0}^{-1}\,.
\end{equation}
Here it is necessary to point out the different level of 
approximation we used in deriving Eqs.~(\ref{meff}) and~(\ref{swfsc}).  
Therefore it is not possible to verify the HM relation~(\ref{meffhomo})
by the above strong--coupling calculation of $Z_{K=0}$.  
The leading exponential dependence of~(\ref{meff}) and~(\ref{swfsc}), 
however, is found to be the same (not the same) for the HM (EHM), 
in good agreement with the numerical results of Figs.~2 and~5.
\bibliography{ref}
\bibliographystyle{phys}
\newpage
\begin{table}
\caption{Quasiparticle weight $Z_{K=0}$ obtained from ED 
($N=8$, $M=24$) and within WCT  
according to Eq.~(\ref{zmahan2}).} 
\begin{tabular}{ccccc}
&\multicolumn{2}{c}{$\lambda =0.1$, $g^2=0.2$}& 
\multicolumn{2}{c}{$\lambda =0.5$, $g^2=1/3$}\\
&ED&WCT&ED&WCT\\
\tableline
HM  & 0.955 & 0.946 & 0.893 & 0.848\\
EHM & 0.918 & 0.893 & 0.857 & 0.781\\
\end{tabular}
\end{table}
\section*{Figure Captions}
Fig. 1: Electron--lattice correlations in the weak-- (a)
and strong--coupling  (b) cases. ED results are obtained for a
finite chain with $N=8$ sites and at most 24 phonons. 
The insets show the weight of the $m$-phonon state in 
the ground state.\\

Fig. 2: Single--particle spectral function $ A_{K}(E)$ (thin lines) 
and partial integrated density of states $N_{K}(E)$ (thick lines)
for the 1D HM (a-b) and EHM (c-d) with $\ho=1.0$ ($N=8$, $M=24$). 
Solid and dot-dashed lines belong to states with total 
momentum $K=0$ and $K=\pi$, respectively.\\

Fig. 3: Band dispersion of the 1D extended Holstein model for low 
(a), intermediate (b), and high (c) phonon frequencies. Exact data 
are extracted from finite-lattice diagonalizations with 
$N=8$ and 10 sites. In the weak-- and strong--coupling regimes ED results 
are compared with the theoretical predictions.\\

Fig. 4: Flattening of the polaron band dispersion in the
1D Holstein model (weak--coupling case). Long-- and short--dashed curves 
denotes the solution of Eq.~(\ref{bdwc}) and the bare phonon frequency 
$\ho =1.0$, respectively. The agreement of ED and WCT gets better 
as $\lambda$ decreases.\\

Fig. 5: Inverse effective polaron mass for the 1D (extended) Holstein model
at $\ho=1.0$.
The QMC data are taken from Ref.~\protect\cite{AK99}. 
The ratio of the effective masses of the EHM and HM 
polarons are displaced in the inset.\\

Fig. 6: Optical absorption in the 1D Holstein and extended Holstein 
models. The regular part of the conductivity $\sigma^{reg}$ (thin lines)
and integrated spectral weight $\cS^{reg}$ (thick lines) are shown in the
weak-- (a) and strong--coupling (b) regimes.\\

Fig. 7: Renormalized kinetic energy ($S^{tot}$) and contribution
of $\sigma^{reg}$ to the f-sum rule ($S^{reg}$) as a function of
EP coupling ($\lambda$) at $\ho=1.0$.
\newpage
\onecolumn
\pagestyle{empty}
\begin{figure}[!htb]\caption{}
\epsfig{file= 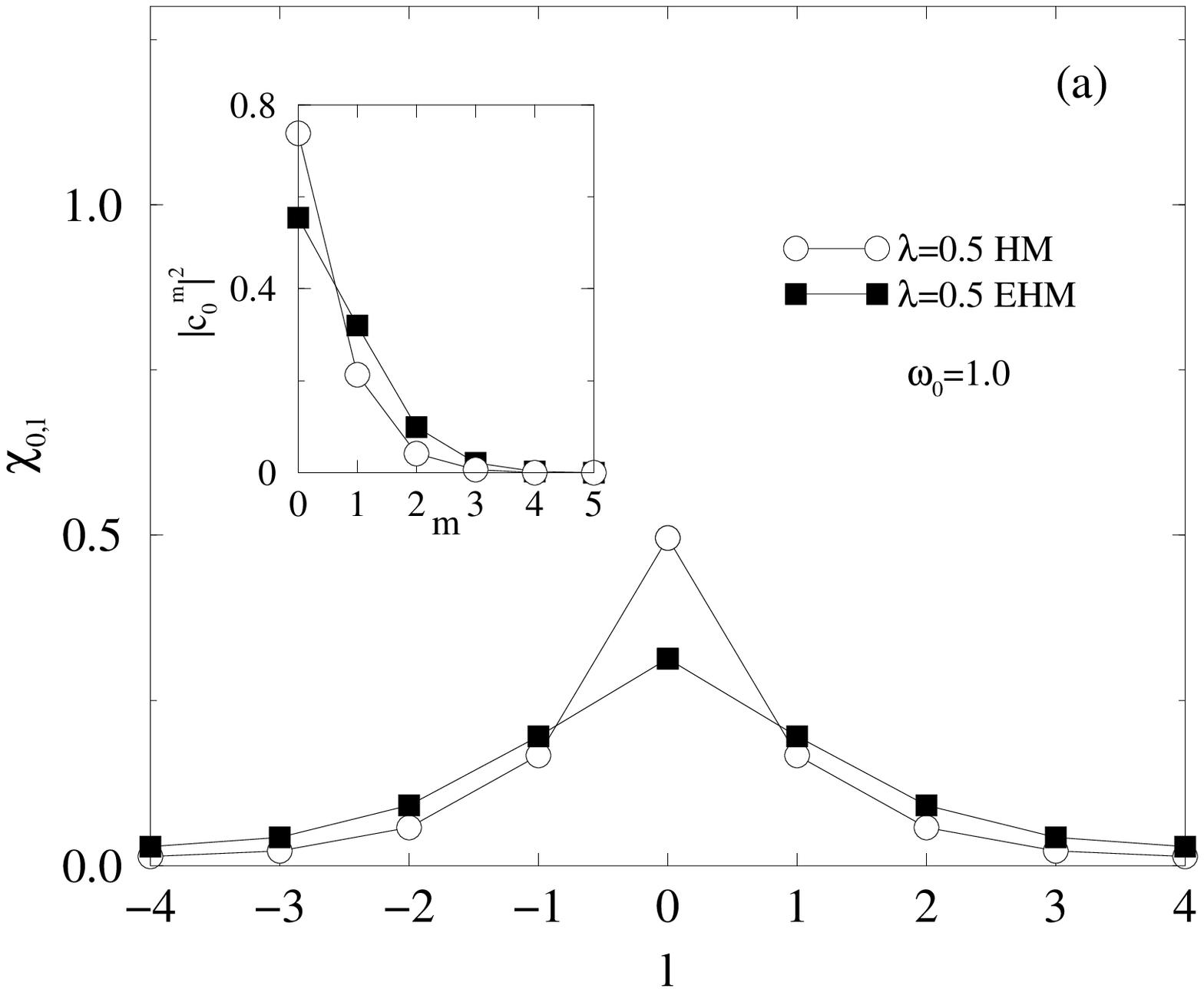, width = 0.8\linewidth}  
\epsfig{file= 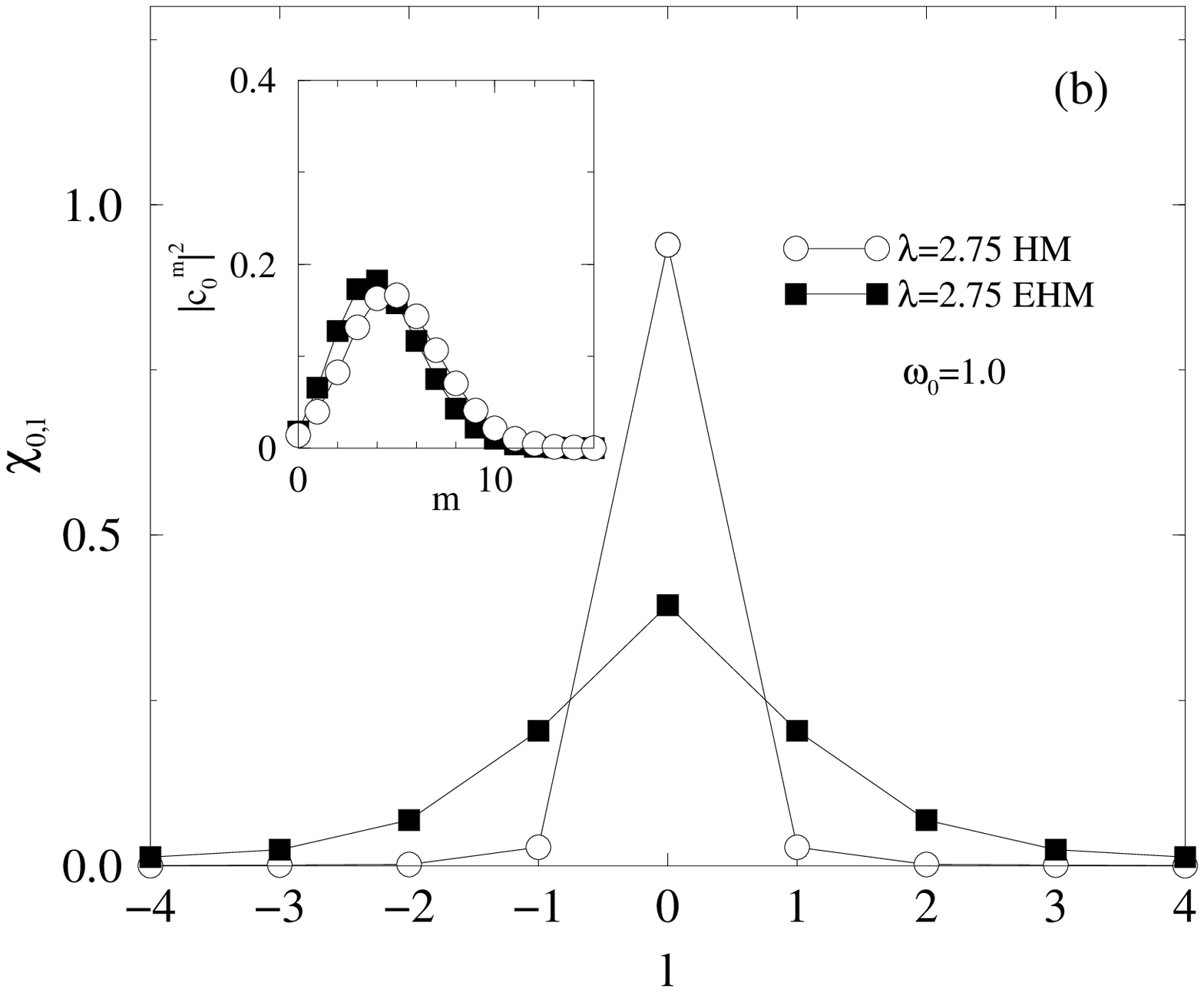, width = 0.8\linewidth}\\
\label{fig1}
\end{figure}
\newpage
\begin{figure}[!htb]
\epsfig{file= 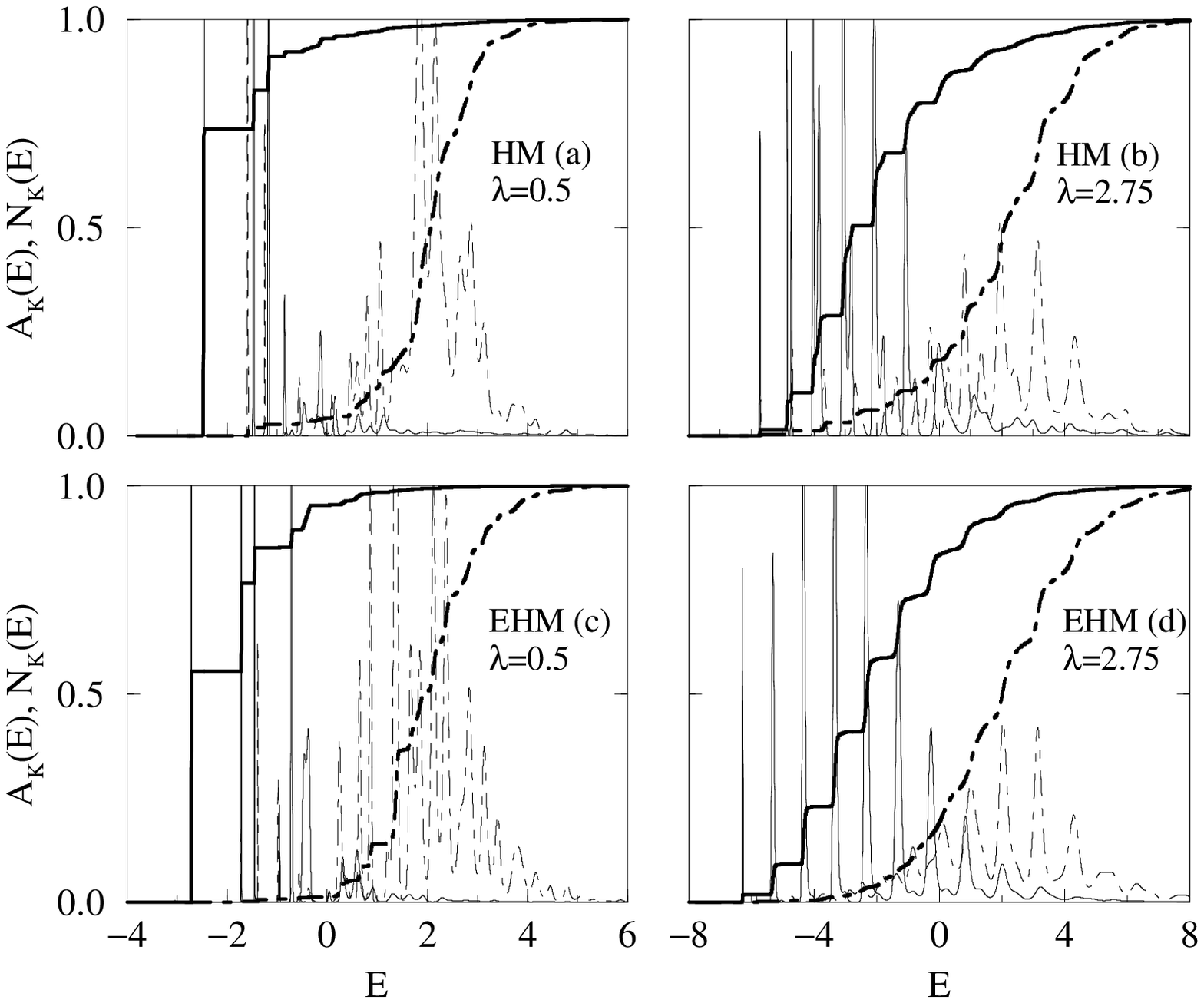, width = \linewidth}\\  
\caption{}
\label{fig2}
\end{figure}
\newpage
\begin{figure}[!htb]
\epsfig{file= 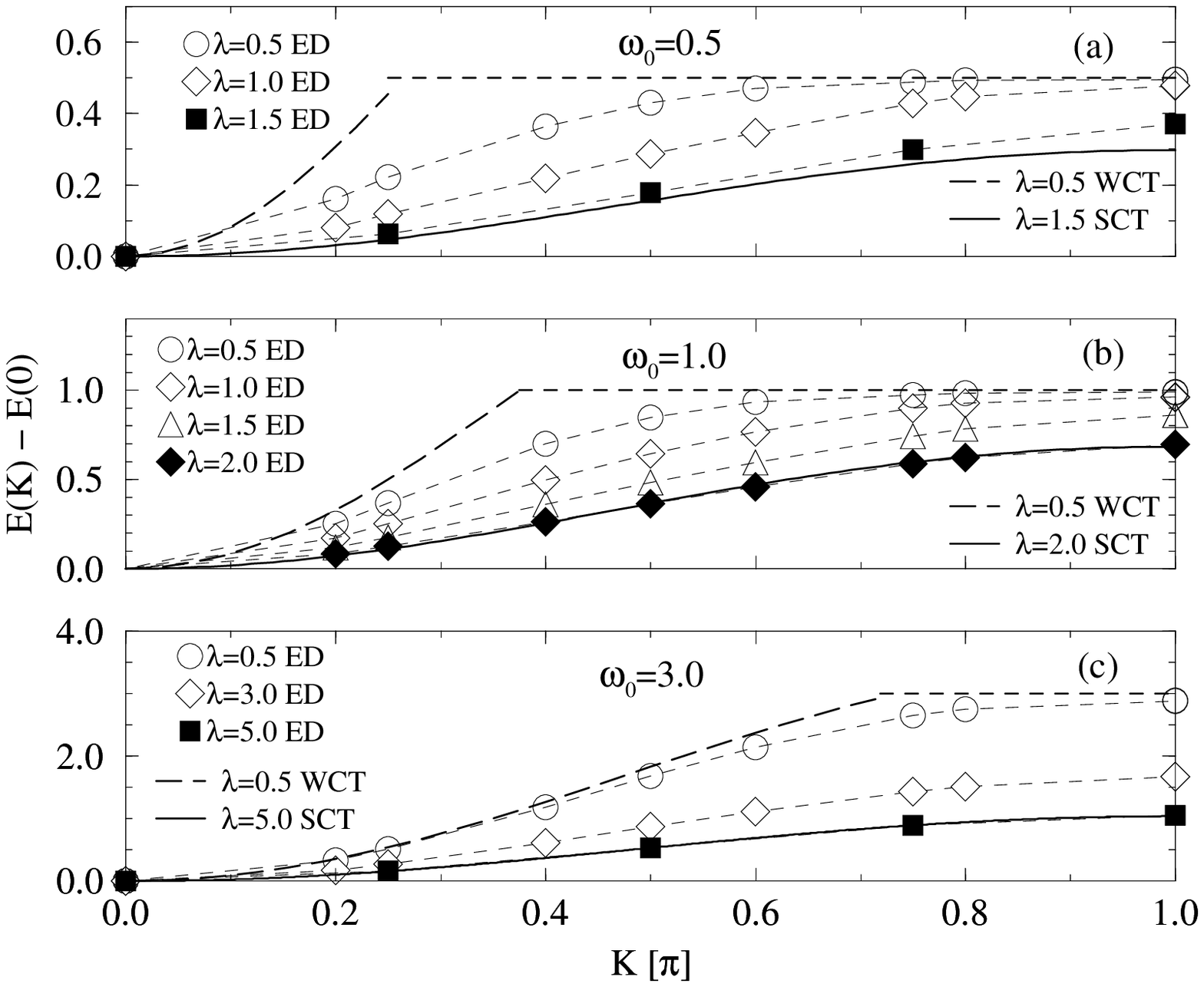 , width = \linewidth}\\  
\caption{}
\label{fig3}
\end{figure}
\newpage
\begin{figure}[!htb]
\epsfig{file= 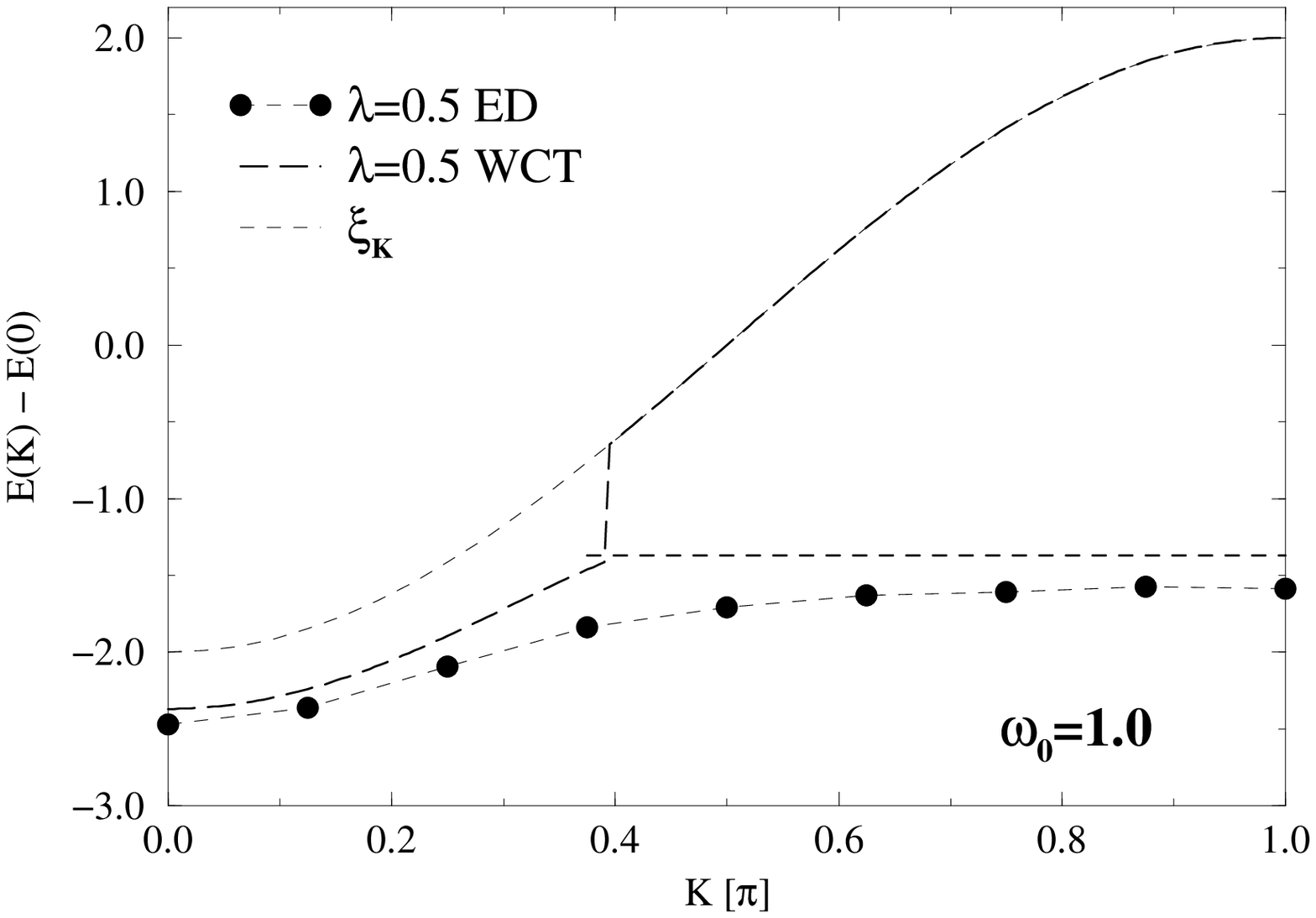 , width = \linewidth}\\ 
\caption{} 
\label{fig4}
\end{figure}
\newpage
\begin{figure}[!htb]
\epsfig{file= 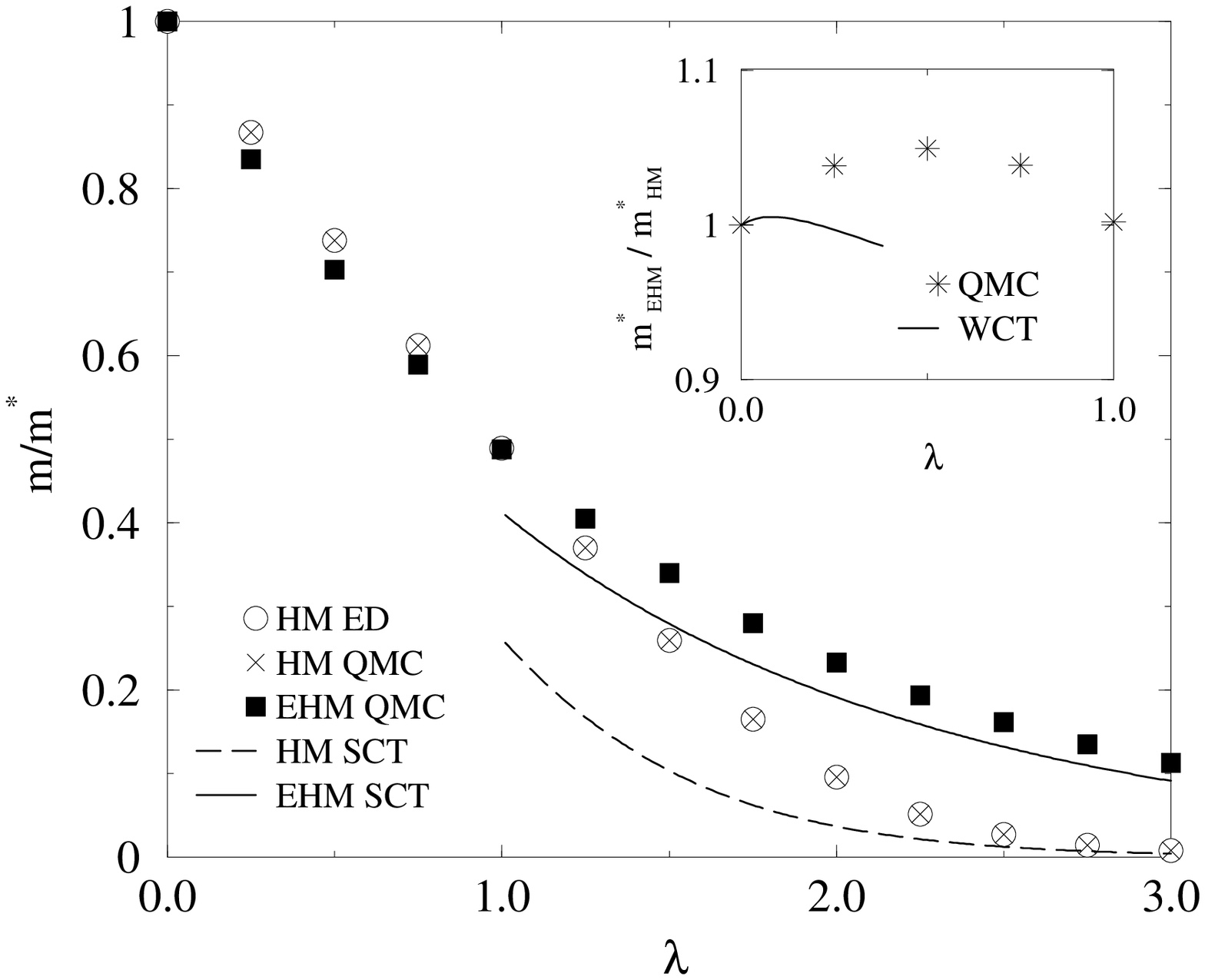, width = \linewidth}\\  
\caption{}
\label{fig5}
\end{figure}
\newpage
\begin{figure}[!htb]
\epsfig{file=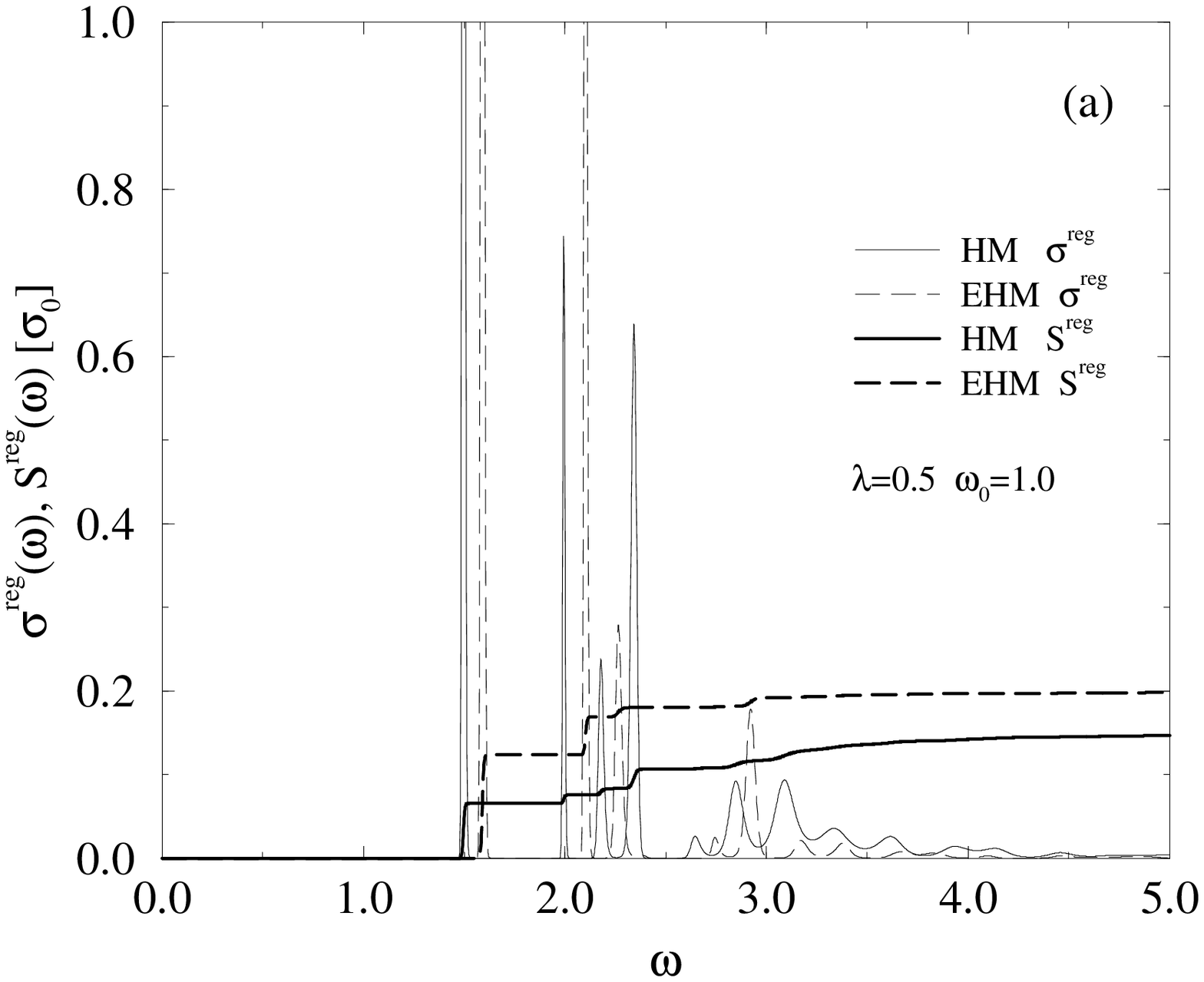 , width = .8\linewidth}  
\epsfig{file=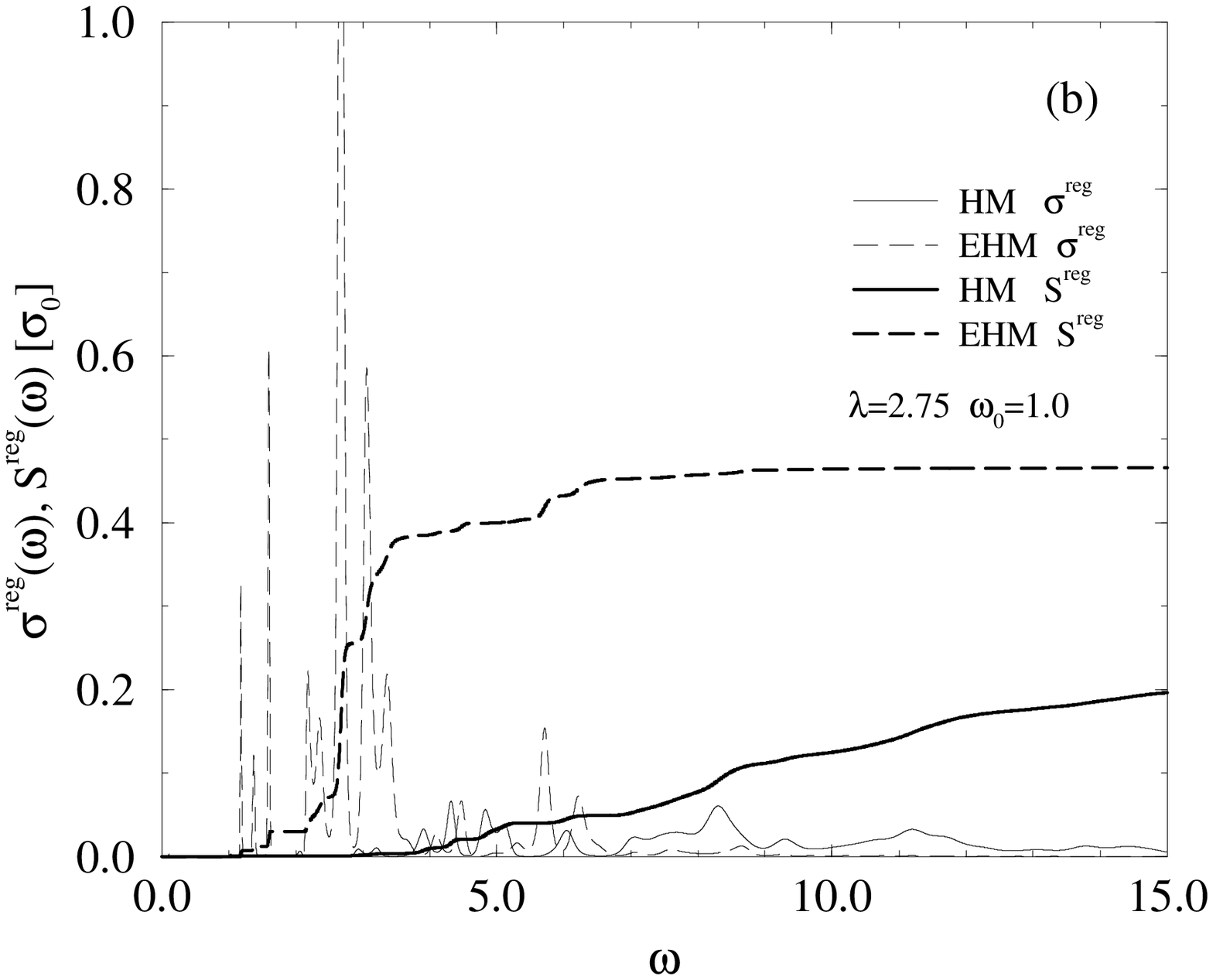 , width = .8\linewidth}\\
\caption{}
\label{fig6}
\end{figure}
\newpage
\begin{figure}[!htb] 
\epsfig{file= 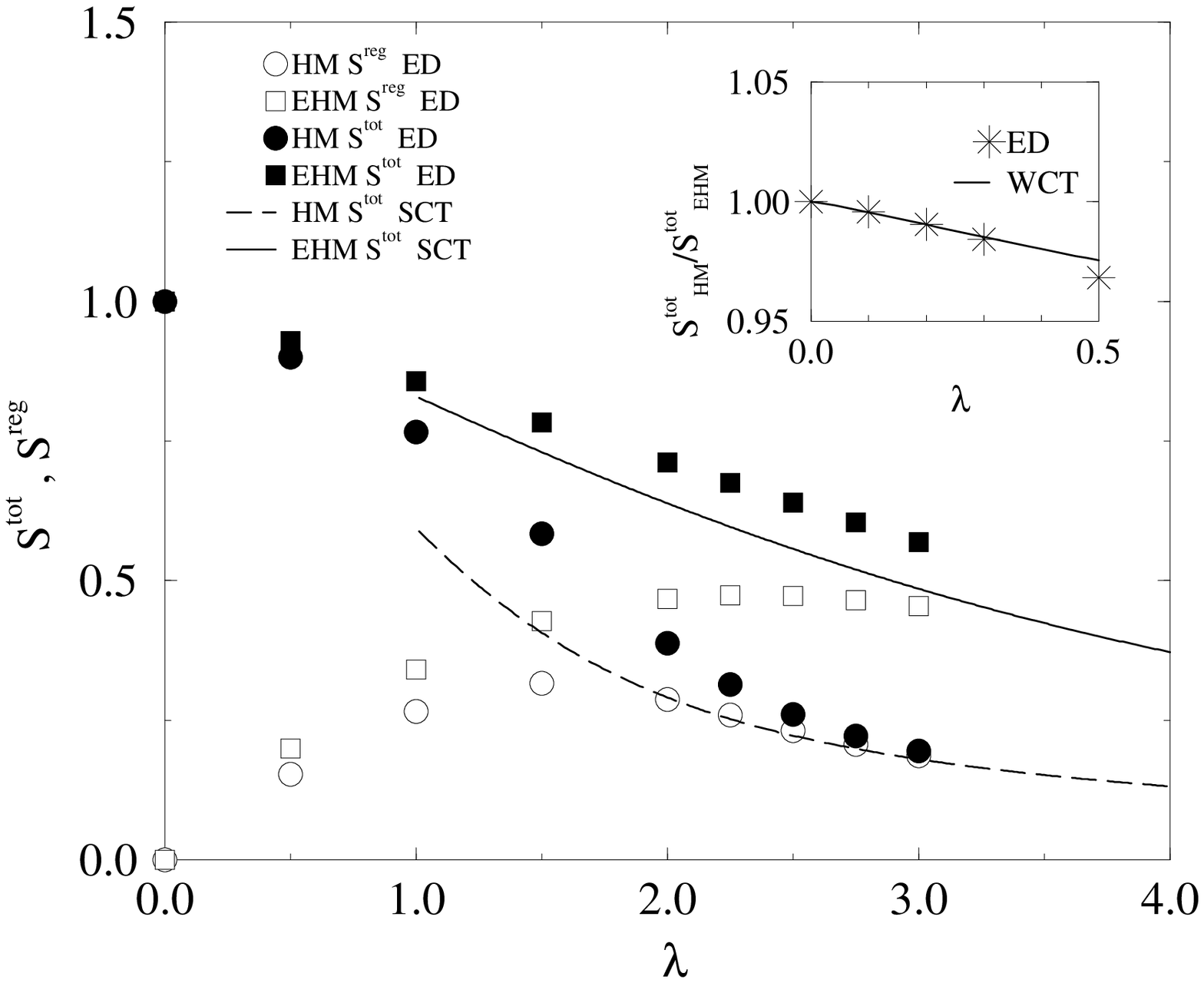, width = \linewidth}\\
\caption{\hfill}
\label{fig7}
\end{figure}
\end{document}